\documentstyle[twoside,12pt,epsf,aps]{revtex}
\relax
\newcommand{\MET}{\mbox{$\protect \raisebox{.3ex}{$\not$}\et$}}

\def\W{{\em W\/ }}
\def\Z0{${\em Z^0\/}$}

\def\r#1 {$^{#1}$}

\newcommand{\et}{{\rm E}_{\scriptscriptstyle\rm T}}

\newcommand{\met}{\mbox{$\protect \raisebox{.3ex}{$\not$}\et \ $}}

\newcommand{\gevcc}{ {\rm GeV/c^2}}

\def\gepsfcentered#1{
  \def\testit{#1}
  \def\lbracket{[}
  \ifx\testit\lbracket
    \let\dofilecmd=\gepsfwithopt
  \else
    \let\dofilecmd=\gepsfnoopt
  \fi
  \dofilecmd}

\def\gepsfnoopt#1{
  \begin{center}
  \leavevmode
  \epsffile{#1}
  \end{center}}

\def\gepsfwithopt#1 #2 #3 #4]#5{
  \begin{center}
  \leavevmode
  \gepsfmaxx=0.94\textwidth
  \epsffile[#1 #2 #3 #4]{#5}
  \end{center}}

\newdimen\gepsfmaxx
\def\epsfsize#1#2{
  \ifnum \epsfxsize=0
    \ifnum \epsfysize=0
      \ifnum #1 > \gepsfmaxx
        \gepsfmaxx
      \else
        #1
      \fi
    \else
      \epsfxsize
    \fi
  \else
    \epsfxsize
  \fi
}

\newcommand{\ppton}{p\bar{p} \rightarrow N}
\newcommand{\ppWN}{p\bar{p} \rightarrow W N}
\begin{document}
\bibliographystyle{unsrt}
\title{Additional studies of the anomalous $l+\MET+{\rm 2,3 \; jet}$
       events observed by CDF}
\maketitle
\font\eightit=cmti8
\def\r#1{\ignorespaces $^{#1}$}
\hfilneg
\begin{sloppypar}
\noindent
G.~Apollinari,\r {2}
 M.~Barone,\r {3}  
W.~Carithers,\r {4} 
T.~Dorigo,\r {6} 
I.~Fiori,\r {6} 
P.~Giromini,\r {3}
 F.~Happacher,\r {3} 
M.~Kruse,\r {1} 
S.~Miscetti,\r {3} 
A.~Parri,\r {3}
F.~Ptohos,\r {3}
 and Y.~Srivastava~\r {5}
\end{sloppypar}
\vskip .026in
\begin{center}
\r {1}  {\eightit Duke University, Durham, North Carolina  27708} \\
\r {2}  {\eightit Fermi National Accelerator Laboratory, Batavia, Illinois 
60510} \\
\r {3} {\eightit Laboratori Nazionali di Frascati, Istituto Nazionale di Fisica
               Nucleare, I-00044 Frascati, Italy} \\
\r {4} {\eightit Ernest Orlando Lawrence Berkeley National Laboratory, 
Berkeley, California 94720} \\
\r{5} {\eightit Northeastern University, Boston, Massachusets 02115} \\
\r {6} {\eightit Universita di Padova, Istituto Nazionale di Fisica 
          Nucleare, Sezione di Padova, I-35131 Padova, Italy} \\
\end{center}

\vspace{0.2em}
\begin{abstract}
   We present additional studies of the kinematics of the anomalous events
   observed by the CDF experiment. These events contain a high-$p_T$ lepton,
   large transverse missing energy ($\MET$), and 2 or 3 high-$E_T$ jets, one
   of which contains both a displaced secondary vertex and a soft lepton.
   Previous articles detailed the selection and the kinematic properties of
   these events. In the present paper, we use several phenomenological
   approaches to model these data and to estimate their production cross
   section.
  \\
  PACS number(s): 13.85.Qk, 13.20.He, 14.80.Ly
\end{abstract}
\section{ Introduction}
\label{s-intro}
 The CDF experiment has observed~\cite{anomal} an excess of events in the 
 $\W+2$ and $\W+3$ jet topologies in which the presumed heavy-flavor jet
 contains a lepton in addition to a secondary vertex. The rate of these 
 events (13 observed) is larger than what is predicted by the simulation
 of known standard model (SM) processes (4.4 $\pm$ 0.6 events expected,
 including single and pair production of top quarks). The kinematical
 properties of these events are not consistent with what is expected if 
 the excess were due to a statistical fluctuation of the SM contributions
 ~\cite{anomal,comb}. This small sample of events has been isolated in a
 detailed examination of the heavy-flavor content of the $\W+$ jet data
 sample collected by the CDF experiment during the 1992-1995 collider run
 at the Fermilab Tevatron. Jets with heavy-flavor hadrons are identified using
 the SECVTX algorithm that reconstructs secondary vertices (SECVTX tags),
 the jet-probability (JPB) algorithm~\cite{jpb}, and the SLT algorithm that 
 identifies leptons produced in the decay of $b$ and $c$-hadrons (SLT tags).
 Jets containing both a SECVTX and SLT tag (supertag) are referred to as
 superjets. The data set, the analysis tools and the Monte Carlo simulations
 used in this study are the same as those described in Sec.~IV of 
 Ref.~\cite{anomal}. The characteristics of the 13 events with a superjet are
 listed in Tables~XVI and~XVII of Ref.~\cite{anomal}.

 We are not aware of any model for new physics which incorporates the 
 production and decay properties necessary to explain all features of the 13
 events with a superjet. In the absence of a suitable theoretical model, in 
 this paper we investigate several hypotheses with the purpose of evaluating
 the production cross section of these anomalous events. For this purpose, we
 use tools, names and ideas developed for specific models, none of which can
 be supported or excluded at the current time because of the limited statistics
 of the data. In Sec.~\ref{sec:s-inter} we investigate the hypothesis that 
 superjets are due to the production and weak decay of a light spin-0
 quark (scalar quark, $\tilde{q}$). In order to estimate the size of the
 anomalous contribution we fit the yields of $W+$ jet events with different
 types of tags with a SM simulation implemented with the pair production of
 a scalar quark and a $b$ quark  with large invariant mass 
 ($\simeq$ 220 $\gevcc$). The fit result shows that the observed tagging rates
 can be reasonably described with this simple addition, and provides a first
 indication of how much the CDF measurement~\cite{ttsec} of the $t\bar{t}$
 production cross section is impacted by the presence of events with a 
 superjet. In Sec.~\ref{sec:w-inter} we use several simulations based upon the
 salient features of the data in order to evaluate the detector acceptance for
 events with a superjet and estimate their production cross section.
 Section~\ref{sec:s-concl} summarizes our conclusions.

\section{Estimate of the size of the anomalous production}
\label{sec:s-inter}
 An interpretation in terms of new physics of the superjet properties, which
 are described in Sec.~IX of Ref.~\cite{anomal}, would require the production
 of a low-mass, strongly interacting object, decaying semileptonically with a
 branching ratio close to 1 and with a lifetime of the order of a picosecond.
 The low mass is required by the fact that the SECVTX and SLT tags are 
 contained in a cone of radius 0.4 in the $\eta-\phi$ space around the jet
 axis. The request that the object is strongly interacting is motivated by the
 fact that several tracks in the superjet point directly to the primary vertex,
 suggesting an emission of gluons during the evolution prior to the decay. The
 large semileptonic branching ratio, finally, is required to explain the large
 fraction of soft lepton tags in jets tagged by SECVTX.
 
 Since there are no experimental limits on the existence of a charge-1/3 
 scalar quark with  mass smaller than 7.4 $\gevcc$~\cite{mssm-can,nappi,carena},
 the supersymmetric partner of the bottom quark is a potential candidate.
 In order to explain the observed rate of supertags, one possible assumption
 is that this scalar quark has three-body decay modes to quarks mediated by
 heavy gauge fermions $\chi$ ($m_{\chi} \geq m_{\tilde{b}}$) and that  
 $\chi \rightarrow l \tilde{\nu}$ is the only allowed decay. As an example,
 in this study we use the ansatz that superjets  are due to the production and
 decay into $l c \tilde{\nu}$ of a scalar quark $\tilde{b}$ with a mass of 
 3.6 $\gevcc$ and a lifetime of 1.0 ps; the scalar neutrino is assumed to be
 massless. Such a scalar quark is not ruled out by the CLEO collaboration
 ~\cite{cleo-sb} if its decay matrix element results in a lepton spectrum 
 softer than three-body phase space, as is the case for the matrix element
 used by us (see Appendix~A). Even if such a $\tilde{b}$ quark were ruled out, 
 we would retain this simple model, which fits the data reasonably well, as a
 working hypothesis to estimate the production cross section of the anomalous
 events. In fact, many slightly different conjectures ($m_{\tilde{\nu}}$ in 
 the range of a few $\gevcc$ and 
 $ 2 \leq m_{\tilde{b}} - m_c - m_{\tilde{\nu}} \leq 5\;\gevcc $) are 
 still not excluded by any data analysis and, because of the limited
 statistics of the CDF data and the large transverse energy of the superjets 
 ($\simeq$ 60 GeV), they also provide  a reasonable modeling of the superjet
 properties~\cite{limit}. We also note that we assume the presence of a $c$ 
 quark in the $\tilde{b}$ decay but this assumption is not required to fit any
 of the superjet properties. 

 The estimate of the production cross section of the events with a superjet
 requires the knowledge of the tagging efficiencies and the detector acceptance,
 which in turn depend on the kinematics of the data. For example, the 
 acceptance is sensitive to the pseudo-rapidity and transverse momentum 
 distributions, which are anomalous in the data. We use three techniques of
 increasing complexity to estimate the sensitivity to the details of
 modeling the kinematics. In the first method, we ignore entirely the 
 kinematics of the primary lepton and the missing energy, and we model the jet
 kinematics  with the fictitious production of a heavy state $N$ of mass
 220 $\gevcc$ which decays into $\tilde{b} \bar{b}$ according to phase space.
 This mechanism is chosen to reproduce the hard $E_T$ spectrum of the superjet
 and the other jets in the final state, as well as their invariant mass 
 distribution. The observed rates of tagged events are fitted with 
 a SM simulation implemented with the additional production
 $p\bar{p}\rightarrow N \rightarrow \tilde{b} \bar{b}$ (described in  
 Appendix A). Since this model is the simplest, we describe the fit in some 
 detail in subsection~A. Using the result of this fit, subsections~B-D 
 show that the addition of this hypothetical process improves
 the simulation of other peculiar features of the events with a superjet.
\subsection{Fit of the tagging rates}
\label{sec:ss-tagrates}
 We fit the rates of $W+ n$ jet events ($n$=1,4) before and after tagging with
 the SM simulation implemented with the $N$ production using a likelihood
 technique. In order to use all the information contained in the data, we fit
 nine ($k=1,9$) different classes of tagged events: events with one and two
 SECVTX tags; events with one and two JPB tags; events with one and two SLT
 tags; events with one supertag; events with one supertag and an additional
 SECVTX tag; and events with one SLT and one SECVTX tag in two different jets.
 In total we fit 31 data points.

 The $t\bar{t}$ cross section and the number of events due to $N$ production
 are unconstrained parameters of the fit, which has  29 degrees of freedom.
 The remaining eleven contributions $X_{j}$ to the tagging rates
 [SECVTX, JPB and SLT mistags, $\W\W$, $\W Z$, $ZZ$, single top, $\W c$, 
 $\W cc$, $\W bb$ and unidentified-$Z$ with heavy flavors] are also fit
 parameters. However, we constrain each contribution to its expected value,
 $\bar{X}_i$, and within its estimated uncertainty, $\sigma_{\bar{X}_i}$, 
 using in the log-likelihood function the term
 \[ G_i=-\frac {(X_i-{\bar{X}_i})^{2} } {2 \sigma_{\bar{X}_i}^{2}}. \]

 The detector acceptance, the luminosity and the values of the SECVTX, JPB 
 and SLT tagging efficiencies are five additional fit parameters, which 
 are also constrained to their measured value within their uncertainties.

 In summary the fit minimizes the function
\[ - \ln {\cal L} = - \sum_{n=1}^{4} \sum_{k=1}^{9} 
     \ln P_{nk}(N_{nk}|\bar{\mu}_{nk}) - \sum_{i=1}^{16} G_i \]
 where $P_{nk}$ is the Poisson probability of observing $N_{nk}$ events when
 $\bar{\mu}_{nk}$ events are expected.

 The minimum value of $- ln {\cal L}$ returned by the fit is -1854.7. 
 If each data point were equal to the best fit result, we would have found 
 $- ln {\cal L}_{0}$ = -1868.7. The difference between $- ln {\cal L}$ and 
 $- ln {\cal L}_{0}$ corresponds to a $\chi^2$ increase of 28 units for 29 
 degrees of freedom. Rates of observed and fitted $W+ \geq$ 1 jet events 
 before and after tagging are compared in Tables~\ref{tab:tab_10.0}~
 to~\ref{tab:tab_10.2}.
 
 The value of the $t\bar{t}$ cross section returned by the fit is 
 4.0 $\pm$ 1.5 pb, in agreement with theoretical expectations~\cite{tsig-pred}
 and the D\O~result~\cite{d0}. This cross section corresponds to the presence
 of 41.9 $\pm$ 15.7 top events in the $\W+ \geq$ 1 jet sample before tagging.
 In the same $\W+ \geq$ 1 jet sample, the fit returns 52.8 $\pm$ 22.1 events 
 due to the production of a massive $\tilde{b}\bar{b}$ pair. We quote numbers
 of events and not a cross section since the lepton and missing energy 
 kinematics have been ignored.
\newpage
\begin{table}[p]
\begin{center}
\caption[]{Composition of the $W+ \geq$ 1 jet sample before tagging.
This composition is determined by a fit to the tagging rates of the data (see text). }
 \begin{tabular}{lcccc}
 Source &  $W+1 \,{\rm jet}$ &  $W+2 \,{\rm jet}$ & $W+3 \,{\rm jet}$ & $W+\geq 4 \,{\rm jet}$ \\ 
\hline
  Data  & 9454 &1370   &198   &  54 \\
  Non-$W$     &  560.0 $\pm$15.1& 71.2 $\pm$ 2.6&  12.3 $\pm$ 2.0 & 5.1 $\pm$ 1.6\\
 $WW,WZ,ZZ$  &  35.8 $\pm$ 7.6&   36.2 $\pm$ 7.7&   6.2 $\pm$ 1.3 & 0.9$\pm$ 0.2\\
 Unidentified-$Z +\rm jets$ & 234.8 $\pm$ 14.5&  38.5$\pm$ 5.9& 7.9 $\pm$
2.4 &   0.7 $\pm$ 0.7\\
 Single top &  14.0 $\pm$ 2.2&    7.8 $\pm$ 1.3 &   1.7 $\pm$ 0.3 & 0.3$\pm$ 0.1\\
$t\bar{t}$ &   1.4 $\pm$ 0.5 &   7.9 $\pm$ 3.0 &  15.9 $\pm$ 6.0 & 16.7$\pm$ 6.3\\
W + jets without h.f. &7955.8$\pm$ 73.6&  1004.5$\pm$ 20.5 & 110.6 $\pm$6.6 & 19.1$\pm$  4.9 \\
$Wc$   &  399.2 $\pm$89.1&   82.1 $\pm$18.3 &   10.0 $\pm$ 2.2 &  1.8$\pm$0.4\\
$Wc\bar{c}$ &  171.6 $\pm$36.9&   60.1 $\pm$12.9 &  10.4 $\pm$ 2.2&  2.2$\pm$0.5  \\
$Wb\bar{b}$ &   67.6 $\pm$ 7.8&   28.8 $\pm$ 3.3 &   5.1 $\pm$ 0.6&  1.4 $\pm$0.2\\
$N$ &   2.2 $\pm$ 0.9&   28.2 $\pm$11.8 &  16.7 $\pm$ 7.0  &  5.6 $\pm$ 2.3\\
\end{tabular}
\label{tab:tab_10.0}
\end{center}
\end{table}

\newpage
\begin{table}[p]
\squeezetable
\begin{center}
\caption[]{Summary of observed  and fitted number of $W$ events with one
(ST) and two (DT) tags of the same kind (SECVTX, JPB, or SLT). 
``Other" is the sum of all processes other than
$t\bar{t}$ and the fictitious $N$ production (see text).}
 \begin{tabular}{lcccc}
  Source &  $W+1 \,{\rm jet}$ &  $W+2 \,{\rm jet}$ &  $W+3 \,{\rm jet}$ &$W+\geq4 \,{\rm jet}$ \\ 
\hline
\hline
 \multicolumn{5}{c}{SECVTX tags} \\ 
 Other (ST)    &  $63.68 \pm 3.89$     & $25.60 \pm 1.64$ & 5.73 $\pm$ 0.39 
& 1.45  $\pm$ 0.13\\
 Other (DT)    &      &$ ~1.56 \pm 0.23$& 0.30 $\pm$ 0.04&  0.07 
 $\pm$ 0.01\\
$t\bar{t}$  (ST)    &    ~0.42 $\pm$ 0.12 &  ~2.59 $\pm$ 0.74 &  5.25 $\pm$ 1.50 &  5.75 $\pm$ 1.65\\
$t\bar{t}$  (DT)     &     &   ~0.58 $\pm$ 0.17  & 2.22 $\pm$ 0.63 &  3.05 $\pm$ 0.87\\
$N$ (ST)             & 	~0.21 $\pm$ 0.06 &  ~9.32 $\pm$ 2.75 &  6.20 $\pm$ 1.83  & 2.01 $\pm$ 0.59\\
$N$ (DT)  		&  &  ~3.08 $\pm$ 0.91 &  1.71 $\pm$ 0.51 &  0.70 $\pm$ 0.21\\
\hline
 Fit (ST)  &	64.30 $\pm$ 3.90 & 37.50 $\pm$ 2.41&  17.17 $\pm$ 1.50 &  9.21 $\pm$ 1.40\\
 Fit (DT)  & ~0.00 $\pm$ 0.00&   ~5.22 $\pm$ 0.81 &  ~4.23 $\pm$ 0.54  & 3.81 $\pm$ 0.77\\
Data  (ST)   & 66&  35 & 10&  11\\
Data  (DT)  &  0&   5 &  6 &  2\\
\hline
\hline
 \multicolumn{5}{c}{JPB tags} \\
 Other (ST)    &   128.46 $\pm$  9.36  & 45.77 $\pm$ 3.46& 9.61 $\pm$ 0.78& 
2.06  $\pm$ 0.16 \\
 Other (DT)    &   & ~1.54 $\pm$ 0.23& ~0.28 $\pm$0.04 &
  0.06 $\pm$0.01 \\
 $t\bar{t}$  (ST)       & ~~0.41 $\pm$ 0.12&   ~2.46 $\pm$ 0.73 &  ~5.12 $\pm$
1.51&   5.46 $\pm$ 1.61\\
 $t\bar{t}$  (DT)       &  &   ~0.56 $\pm$ 0.17  & ~2.02 $\pm$
0.60&   2.81 $\pm$ 0.83\\
 $N$ (ST)               & ~~0.24 $\pm$ 0.08&   ~9.60 $\pm$ 2.97 &  ~5.70 $\pm$ 1.76
&  1.79 $\pm$ 0.55\\
 $N$ (DT)               &  &   ~3.20 $\pm$ 1.00 &  ~1.93 $\pm$
0.60&   0.75 $\pm$ 0.23\\
\hline
 Fit (ST)        & 129.12 $\pm$ 9.44 & 57.83 $\pm$ 5.20 & 20.43 $\pm$
2.66 &  9.31 $\pm$ 1.82\\
 Fit (DT)        &  &   ~5.30 $\pm$ 0.94  & ~4.23 $\pm$ 0.63 
& 3.62 $\pm$ 0.76\\
Data  (ST)      & 125 & 62&  21 & 12\\
Data  (DT)       &  &  6  & 5 &  3\\
\hline
\hline
 \multicolumn{5}{c}{SLT tags} \\
 Other (ST)    &   140.55 $\pm$ 7.47  & 45.41 $\pm$ 2.38 & 10.13 $\pm$ 0.55&
 3.72  $\pm$ 0.22\\
 Other (DT)    &    & ~0.08 $\pm$ 0.01 & ~0.01 $\pm$ 0.00 &
  0.00 $\pm$ 0.00 \\
$t\bar{t}$  (ST)& ~~0.10 $\pm$ 0.03&   ~1.02 $\pm$ 0.32 &  ~2.16 $\pm$ 0.68 &
  2.55
$\pm$ 0.80\\
$t\bar{t}$  (DT)&  &   ~0.03 $\pm$ 0.01 &  ~0.10 $\pm$ 0.03 &
 0.13$\pm$ 0.04\\
$N$ (ST)& ~~0.10 $\pm$ 0.03 &  ~8.07 $\pm$ 2.43 &  ~5.20 $\pm$ 1.56  & 1.75 $
\pm$ 0.53\\
$N$ (DT) &  &  ~0.74 $\pm$ 0.22  & ~0.41 $\pm$ 0.12 &  0.19 $\pm$
0.06\\
\hline
 Fit (ST) & 140.75 $\pm$ 7.47 & 54.51 $\pm$ 2.99 & 17.49 $\pm$ 1.44 & 
8.02 $\pm$ 0.77\\
 Fit (DT) &  &  ~0.86 $\pm$ 0.32 &  ~0.52 $\pm$ 0.17 &  0.32
$\pm$ 0.07\\
Data  (ST)&146&  56 & 17  & 8\\
Data  (DT)&   &  0 &  0 &  0\\
 \end{tabular}
\label{tab:tab_10.1}
\end{center}
\end{table}
\widetext

\newpage

\begin{table}[p]
\begin{center}
\caption[]{Summary of observed  and fitted number of $W$ events with one 
supertag  (ST) and one supertag with an additional 
SECVTX tag (DT) (top table) and  with one SLT
           and one SECVTX tag in two different jets (bottom table).}
 \begin{tabular}{lcccc}
  Source &  $W+1 \,{\rm jet}$ &  $W+2 \,{\rm jet}$ &  $W+3 \,{\rm jet}$ &$W+\geq4 \,{\rm jet}$ \\ 
\hline
 Other (ST) &  3.77 $\pm$ 0.35 &   1.93 $\pm$ 0.18   &  0.54 $\pm$ 0.05  & 
 0.05 $\pm$ 0.00 \\
 Other (DT) &     &   0.16 $\pm$ 0.02 &   0.03 $\pm$ 0.00   &
  0.00 $\pm$ 0.00\\
$t\bar{t}$  (ST)   & 0.03 $\pm$ 0.00  &  0.29 $\pm$ 0.03  &  0.57 $\pm$ 0.05  &  0.70 $\pm$ 0.07\\
$t\bar{t}$  (DT)   &   &   0.07 $\pm$ 0.01  &  0.24 $\pm$ 0.02  &  0.37 $\pm$ 0.03\\
$N$ (ST)  &  0.02 $\pm$ 0.00  &  1.54 $\pm$ 0.14  &  1.02 $\pm$ 0.10  &  0.28 $\pm$ 0.03\\
$N$ (DT)  &    &  1.08 $\pm$ 0.10   & 0.63 $\pm$ 0.06  &  0.28 $\pm$ 0.03\\
\hline
 Fit (ST)&  3.82 $\pm$ 0.36  &  3.76 $\pm$ 0.35  &  2.13 $\pm$ 0.20  &  1.03 $\pm$ 0.10\\
 Fit (DT)&   &   1.31 $\pm$ 0.12  &  0.90 $\pm$ 0.08  &  0.65 $\pm$ 0.06\\
Data (ST) &  1   & 6   & 2   & 2\\
Data  (DT)  &    &  2   & 3   & 0\\
\hline
\hline
  Source &  $W+1 \,{\rm jet}$ &  $W+2 \,{\rm jet}$ &  $W+3 \,{\rm jet}$
&$W+\geq4 \,{\rm jet}$ \\
\hline
 Other  &   &   1.10 $\pm$ 0.10   &  0.44 $\pm$ 0.04  &  
0.20 $\pm$ 0.02 \\
  $t\bar{t}$  &  &  0.22 $\pm$ 0.02 &  0.60 $\pm$ 0.04 &  0.92$\pm$ 0.06\\
   $N$       &  &  1.17 $\pm$ 0.11 &  0.89 $\pm$ 0.08 &  0.32$\pm$ 0.03\\
\hline
 Fit  &  &  2.49 $\pm$ 0.23 &  1.93 $\pm$ 0.18 &  1.44
$\pm$ 0.14\\
  Data        &  &  1 &  0 &  1\\
 \end{tabular}
\label{tab:tab_10.2}
\end{center}
\end{table}

%
\subsection{Choice of the $N$ mass}
~\label{sec:bsbmass}
 The additional production $N \rightarrow \tilde{b}\bar{b}$ is a simple model
 capable of reproducing the observed yield of tagged events as a function of
 the jet multiplicity (see Tables~\ref{tab:tab_10.1} and~\ref{tab:tab_10.2}).
 Since the tagging rates are not very sensitive to the $N$ mass, we have 
 chosen its value after comparing the invariant mass distribution of 
 jet pairs in the data and in the simulation. For this purpose, we have added
 to the original 13 events the four additional events with a superjet listed
 in Tables~XVIII and~XIX of Ref.~\cite{anomal}. Of these 17 events, 11 events
 contain only two jets. In the remaining 6 events with three jets, when 
 possible, we combine the superjet with the other jet tagged by SECVTX 
 (3 events). If no additional jets are tagged, we select the one with the 
 highest transverse energy (3 events). In simulated events due to $N$ 
 production containing two untagged jets in addition to the superjet, the
 highest-$E_T$ jet is produced by a parton from the $N$-decay in 77\% of the
 cases.
  
\begin{figure}[htb]
\begin{center}
\leavevmode
\epsffile{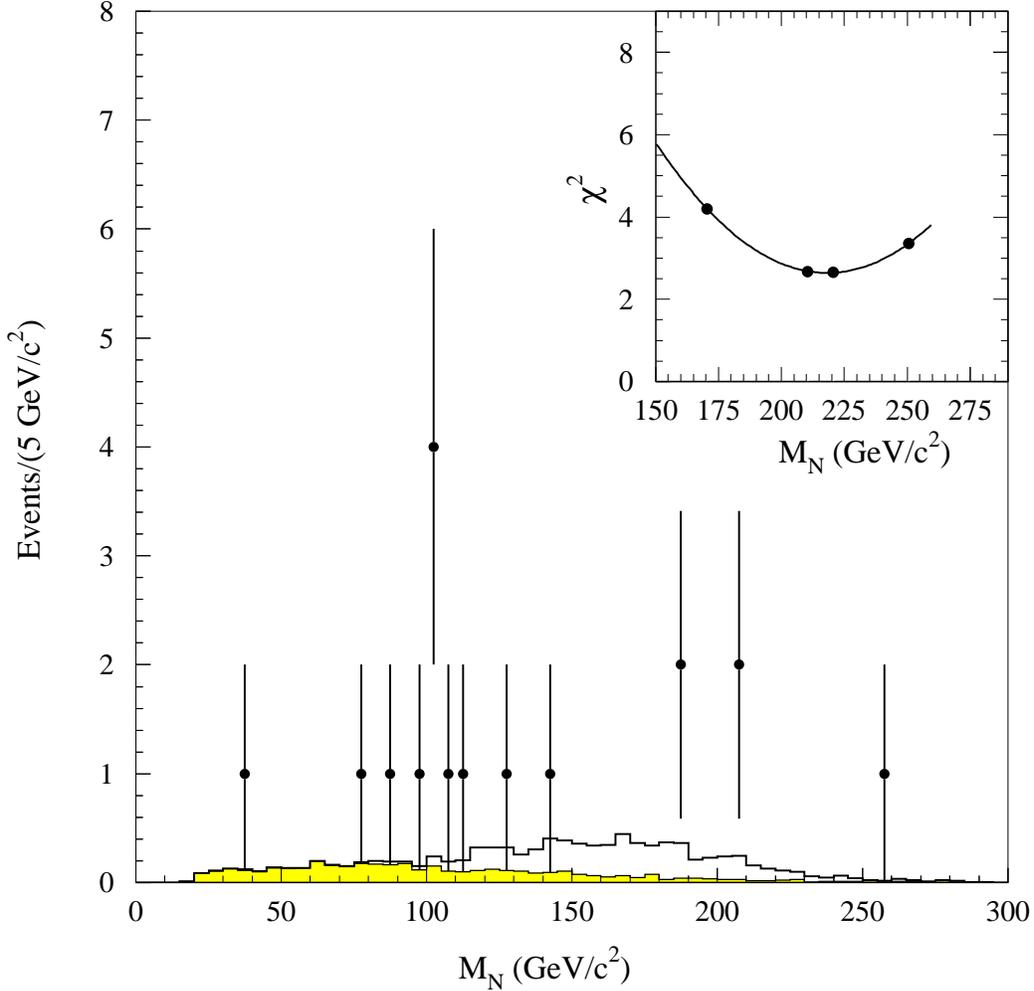}
\caption[]{Di-jet invariant mass distribution in the 17 events with a 
           superjet ($\bullet$) and in the simulation implemented with the 
           production of a state $N \rightarrow \tilde{b}\bar{b}$ with a 
           220 $\gevcc$ mass (histogram); the SM contribution is represented
           by the shaded histogram. The inset shows the yield of the reduced
           $\chi^{2}$ of the comparison between the di-jet invariant mass 
           distributions in the data and in the simulation as function 
           of $M_N$.}
\label{fig:fig_10.2a}
\end{center}
\end{figure}

 Using the normalization provided by the fit in Section~\ref{sec:ss-tagrates},
 the simulation predicts 10.2 events with a supertag: 6.0 $\pm$ 0.6 are due 
 to $N$ production, 1.3 $\pm$ 0.1 are due to $t\bar{t}$ production and 
 2.9 $\pm$ 0.3 are due to the remaining processes. Figure~\ref{fig:fig_10.2a} 
 compares the observed and predicted invariant mass distributions. 
 A Kolmogorov-Smirnov (K-S) test of these distributions, described in detail in Section~VII~C of 
 Ref.~\cite{anomal},  yields a distance $\delta^0$ = 0.37 and a probability
 $P$ = 10.3\%. The result of the K-S comparison of the di-jet invariant mass
 distributions in the data and in the simulation for several $M_N$ values is
 also shown in Figure~\ref{fig:fig_10.2a} (the probability $P$ of the K-S test
 has been converted into a $\chi^{2}$ per degree of freedom). In conclusion,
 the kinematics of the data is compatible with the additional production of a 
 pair of $\tilde{b}\bar{b}$ partons with an average invariant mass of about
 220 $\gevcc$.
%
\subsection{Transverse momentum of the soft lepton tags}
\begin{figure}[htb]
\begin{center}
\leavevmode
\epsffile{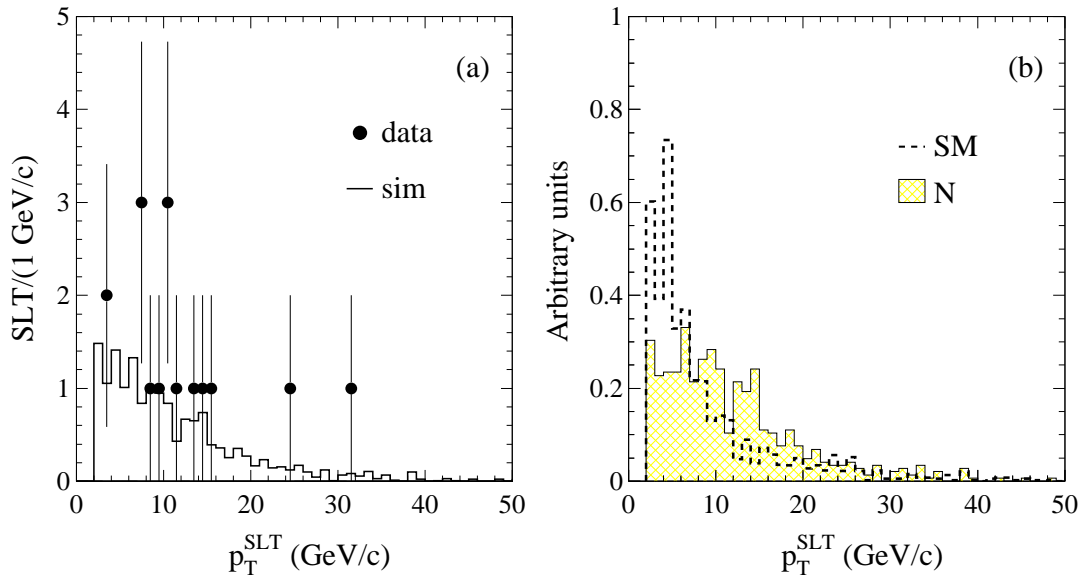}
\caption[]{Distributions (a) of the transverse momentum of soft lepton tags
           in the 15 events with a superjet and in  simulated events which
           include the $N \rightarrow\tilde{b}\bar{b}$ production. The SM and
           the $N$ contributions, normalized to the same number of events, 
           are also shown (b) separately.}
\label{fig:fig_10.3}
\end{center}
\end{figure}

 As discussed in Section~IX~B of Ref.~\cite{anomal}, the distribution of 
 the transverse momentum of the soft lepton tag, $p_T^{SLT}$, in the original
 13 events with a superjet is quite anomalous, and has a $P$ = 0.09\% 
 probability of being consistent with the SM prediction.
 Figure~\ref{fig:fig_10.3}(a) shows the $p_T^{SLT}$ distribution
 in 15 events with a superjet (we add the two events recovered using
 primary  plug electrons listed in Table~XIX of Ref.~\cite{anomal} but not
 the two events triggered by the soft muons because of the trigger 
 $p_T$-cut). The probability that the $p_T^{SLT}$ distribution for 
 these 15 events is consistent with the SM simulation remains small 
 ($P$ = 0.1\%). The  K-S comparison of the  $p_T^{SLT}$ distribution to a SM 
 simulation implemented with the $N$ production (Figure~\ref{fig:fig_10.3}(a))
 yields a distance $\delta^0$ = 0.34 and a probability $P$ = 11.8\%.
 The improved agreement obtained by adding the $N$-production
 can be understood from the comparison of the $p_T^{SLT}$ distributions
 in SM processes and in events due to the $N$ production in which most of 
 the soft leptons are produced by $\tilde{b}$ decays
 (Figure~\ref{fig:fig_10.3}(b)).
%
\subsection{Charge correlation between the primary lepton
            and the soft lepton in the superjets}
 In   13 out of 17 events with a superjet, listed in Tables~XVI to XIX 
 of Ref.~\cite{anomal}, the charges of the primary lepton and the soft lepton
 tag are opposite. SM processes which contribute to events with a superjet do 
 not produce a visible charge correlation; the binomial probability of an equal
 or larger statistical fluctuation is 2.4\%.

 In the $N \rightarrow \tilde{b} \bar{b}$ simulation most superjets are 
 produced by semileptonic $\tilde{b}$-decays. In 67\% of the cases, the soft 
 lepton charge has the same sign of the $\tilde{b}$ charge. According to the
 fit in Section~\ref{sec:ss-tagrates}, 4.2 of the 17 events are attributed
 to SM processes which do not produce a visible charge correlation. If the 
 remaining 12.8 events with a superjet represent the final state 
 $ (\tilde{b} \bar{b}) + l^{+} +\MET$  (and its charge-conjugate), in which 
 the charges of the primary lepton and the $\tilde{b}$ quark are opposite, 
 then in 17 events we expect to find 10.7 events in which the primary and 
 soft lepton have opposite charge. The probability that the data are 
 generated according to this hypothesis increases from 2.4\% to 18.6\%. 
 \section{Estimate of the acceptance for events with a superjet}
 \label{sec:w-inter}
 The fit in the Section~\ref{sec:ss-tagrates} returns 52.8 $\pm$ 22.1 events
 not attributed to SM processes. In order to turn this number into a cross 
 section we try two phenomenological approaches to estimate the detector 
 acceptance. We first use a $\W$+ Higgs boson simulation where we 
 identify the Higgs with the state $N$ which decays to $\tilde{b}\bar{b}$.
 As detailed in Appendix~B, we arbitrarily weight the angular distribution 
 of the primary lepton to obtain an approximate agreement with the data. 
 Although, as shown in Appendix B, the remaining kinematics is poorly modeled
 by the hypothesis of a final state containing two massive particles ($\W$ and
 $N$), this simulation provides a jet multiplicity distribution consistent 
 with the simpler $N$  production. A fit of the rates of tagged $\W+$ jet 
 events with the SM simulation implemented with this $WN$ production also 
 returns a number of tagged $WN$ events consistent with the fit shown in
 Tables~\ref{tab:tab_10.1} and~\ref{tab:tab_10.2}. The detector acceptance 
 for such a $WN$ process is about 9\% without including the $\W$ leptonic
 branching ratio.

 In a second attempt to evaluate the detector acceptance, we model the 
 production of events with a superjet with a more general $2 \rightarrow 4$ 
 hard scattering process. As detailed in Appendix C, the matrix element of 
 this process is tuned using an effective Lagrangian approach to account for
 the salient kinematic features of the data.
 In this picture  the primary lepton and the missing energy do not come from
 $\W$ decays. In this simulation, the yield of events before and after tagging 
 as a function of the jet multiplicity is similar to that of
 the $p\bar{p} \rightarrow N$ simulation (Tables~\ref{tab:tab_10.0}
 to \ref{tab:tab_10.2}). Under the assumption that the lepton in the final state
 can be an electron, a muon or a $\tau$ with equal probability, the detector 
 acceptance for this simulated process is 10.8\%, consistent with the $WN$
 simulation.

 If we average the two techniques for estimating the detector acceptance,
 then the 52.8 $\pm$ 22.1 events with a superjet required by the fit in 
 Table~\ref{tab:tab_10.0} correspond to a cross section of about 5 $\pm$ 2 pb.
%
 \section{Conclusions}
 \label{sec:s-concl}
 We present additional studies of the kinematics of the anomalous events 
 with a superjet observed by CDF. We model 
 several properties of the superjets with the hypothesis
 of the production and decay of a light scalar quark. Using phenomenological 
 approaches, based on this ansatz, to evaluate the detector acceptance and 
 the tagging efficiencies for events with a superjet, we estimate that their 
 production cross section is approximately 5 $\pm$ 2 pb, independent of the
 details of the kinematic distributions.  At the same time, we derive a new
 estimate of the $t\bar{t}$ cross section value, $ 4.0 \pm 1.5$ pb, using 
 $\W+\geq$ 1 jet events with SECVTX, JPB and SLT tags. This result agrees 
 with the D\O~measurement~\cite{d0}, $\sigma_{t\bar{t}} = 4.1 \pm 2.0$ pb, 
 derived in the same topological channel by using kinematical cuts designed 
 to select $t\bar{t}$ events.
\acknowledgments
 We thank the Fermilab staff and the CDF collaboration for their contributions.
 We are grateful to Stefano~Moretti for implementing new processes into the
 {\sc herwig} generator and to H.~Georgi, S.~Glashow, H.~Logan, M.~Mangano,
 and J.~Terning for many interesting and helpful conversations. This work was
 supported by the U.S.~Department of Energy, the National Science Foundation
 and the Istituto Nazionale di Fisica Nucleare.

\appendix
\section{Simulation of the process $\ppton$}
 We simulate the process $p\bar{p} \rightarrow N$ with the {\sc herwig}
 Monte Carlo program~\cite{herwig}. We produce the state $N$ using option 1605,
 which calculates the production of a Higgs boson in $p\bar{p}$ interactions. 
 The Higgs boson is forced to decay to $b\bar{b}$. The decay of the 
 $\bar{b}$-hadron formed by {\sc herwig} is performed using the {\sc qq} Monte
 Carlo program~\cite{cleo}.
 The $b$-quark is transformed into a scalar quark in the following way.
 Having modified the {\sc herwig} generator, we set the $b$-quark mass to 3.6
 $\gevcc$.  The $b$-quark is hadronized by {\sc herwig} as a fermion.
 At the end of the hadronization process, when a $b$-hadron has been formed,
 we change its identity to a fictitious $\tilde{b}$-hadron, but we keep the
 hadron mass evaluated by {\sc herwig} at this stage. The decay of the 
 $\tilde{b}$-hadron is modeled with {\sc herwig} using the spectator model 
 (routine {\sc hwdhvy}). In this model, {\sc herwig} weights in the routine
 {\sc hwdhqk} the phase space of the three-body semileptonic decay of a 
 $b$-quark with the V-A matrix element calculated with the routine {\sc hwdwwt}.
 In the case of a scalar quark we modified {\sc herwig} to weight the 
 phase-space with the matrix element:
 \begin{eqnarray}
 \frac{d\Gamma}{dz_c dz_l}= K
        [(1-z_c)(1-z_l)-R_{\tilde{\nu}}+R_c(z_c-z_l+R_{\tilde{\nu}}-R_c)]
 \end{eqnarray}
  where $K$ is a normalization factor, $R_c=m_c^2 / m_{\tilde{b}}^2$ and
  $R_{\tilde{\nu}}=m_{\tilde{\nu}}^2 / m_{\tilde{b}}^2$.

\begin{figure}[htb]
\begin{center}
\leavevmode
\epsffile{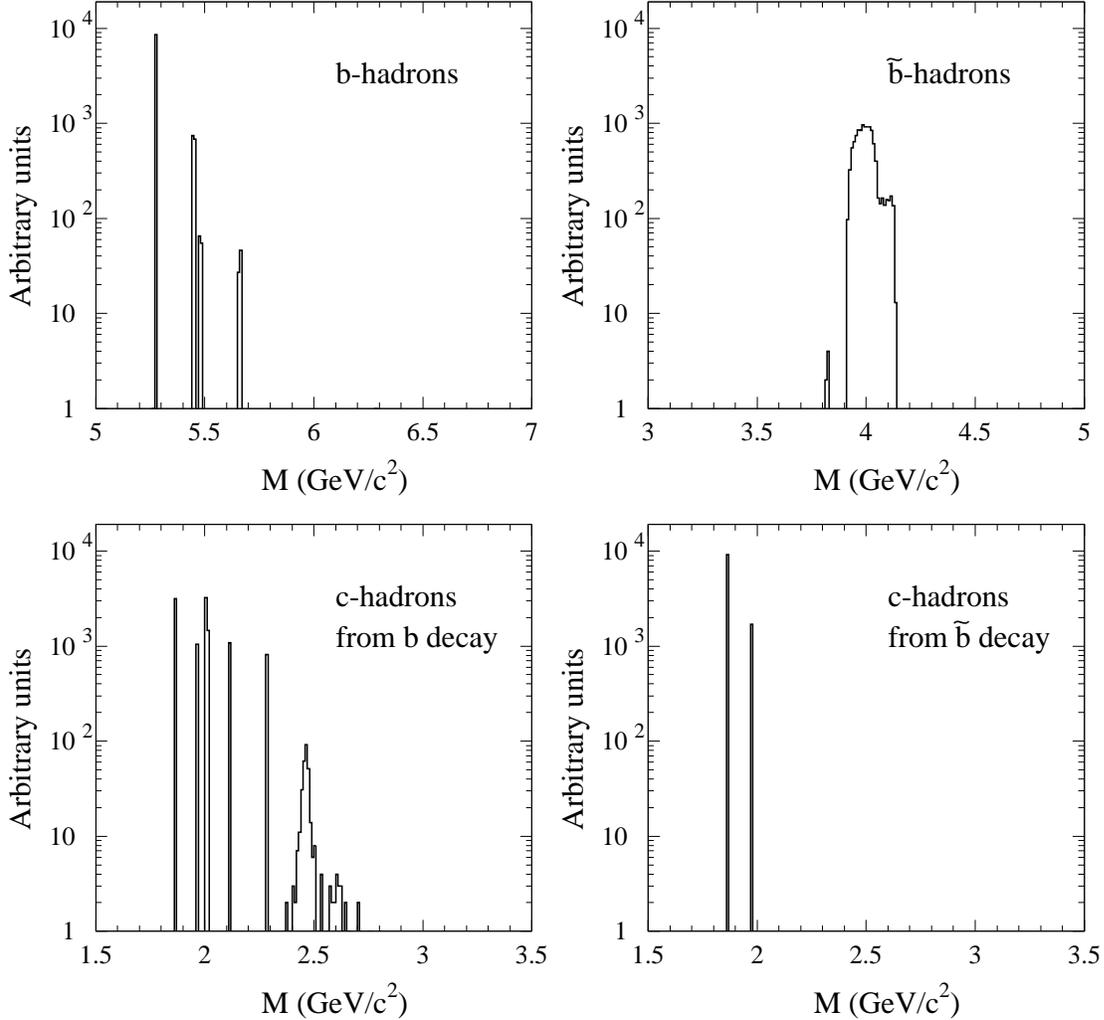}
\caption[]{ Mass spectrum of the $b$ and $\tilde{b}$-hadrons produced by
            the {\sc herwig} simulation using $m_b$=4.75 $\gevcc$ (a)
            and $m_{\tilde{b}}$=3.65 $\gevcc$ (b).
            (c): mass spectrum of the $c$-hadrons from $b$-hadron decays
            modeled with {\sc qq}.
            (d): mass spectrum of the $c$-hadrons from $\tilde{b}$-hadron
            decays modeled with the {\sc herwig} spectator model.}
\label{fig:fig_10.1}
\end{center}
\end{figure}

  Here $z_c$ and $z_l$ are defined as
  \[ z_c = \frac{2p_{\tilde{b}} \cdot p_c}{m_{\tilde{b}}^2}
   \mbox{~~ and~~~ }
   z_l = \frac{2p_{\tilde{b}} \cdot p_l}{m_{\tilde{b}}^2} \]
  The phase space limits are:
        \[ 2 \sqrt{R_c} < z_c < 1 + R_c - R_{\tilde{\nu}} \]
        \[\frac{1+R_c-R_{\tilde{\nu}}-z_c}{1-[z_c-\sqrt{z_c^2-4R_c}]/2} < z_l <
        \frac{1+R_c-R_{\tilde{\nu}}-z_c}{1-[z_c+\sqrt{z_c^2-4R_c}]/2} \]
 This matrix element for the decay $\tilde{b} \rightarrow c l \tilde{\nu}$
 is derived from the tree level calculation outlined in Ref.~\cite{sq-koba}
\scriptsize
\begin{eqnarray}
 {\cal M}= \sum_{j=1}^{2} \frac {-g^{2} V_{j1} F_{\rm L} |V_{c\tilde{b}}| }
 {  (p_{\tilde{b}}-p_c)^{2}-
 m_{\chi_{j}}^{2} }  \left[ \left[  U_{j1} cos \theta - \frac {m_{b}
 U_{j2}
 sin \theta}{\sqrt{2} m_{W} \cos \beta} \right] m_{\chi_{j}} \bar{u} (p_c)
 P_R v(p_l) +\frac {m_c V_{j2}^{*} cos \theta} {\sqrt{2} m_{W} \sin \beta}
 \bar{u}(p_c) {\not\!p_{\tilde{\nu}} } P_R v(p_l)   \right]
\end{eqnarray}
\normalsize
  where $\chi_{j}$'s  are the chargino mass eigenstates, and $U$ and $V^{*}$
  are the mixing matrices for the right and left-handed charginos, respectively.
  The first subscript of $U$, $V$ corresponds to mass eigenstates and the
  second to weak eigenstates (1 for the gaugino and 2 for the higgsino).
  Here tan$\beta$ is the ratio of the vacuum expectation values of the
  two Higgs fields, $\theta$ is the mixing angle between left-handed and
  right-handed scalar quarks,  $F_{\rm L}$ is the fraction of left-handed
  component of the scalar neutrino, and $|V_{c\tilde{b}}|$ is the CKM matrix
  element.  Equation (A1) follows from (A2) when the decay is mediated by the
  higgsino coupling to the right-handed matter. If the decay is mediated by
  the gaugino coupling to the left-handed matter the matrix element is
 \begin{eqnarray}
  \frac {d\Gamma}{dz_c dz_l}= K (z_c+z_l -1+R_{\tilde{\nu}}-R_c ).
 \end{eqnarray}
  In the latter case the two fermions in the final state are both left-handed
  and tend to be produced back-to-back since the initial state is spinless.
  Using equation (A1) the two fermions in the final state have opposite 
  handedness and tend to be produced more collinear than when using equation 
  (A3).  It follows that the matrix element (A1) produces leptons with a 
  momentum distribution appreciably softer than when using the matrix element
  (A3) or a phase space decay, as done in Ref.~\cite{cleo-sb}.  The $c$-quark
  emerging from the $\tilde{b}$-quark decay is then recombined in {\sc herwig}
  with the spectator quark by the routine {\sc hwcfor}. The decay of the 
  resulting $c$-hadron is performed with the {\sc qq} Monte Carlo program.
  In the spectator model, excited $D$-meson states are produced only by the
  hadronic current carried by the virtual $\W$ (so called {\em upper vertex}).
  Since we impose that the gauge fermion involved in the scalar quark decay
  has only leptonic decay modes, the simulation produces a very simple list of
  $c$-hadrons with respect to the {\sc qq} generator (see 
  Figure~\ref{fig:fig_10.1}).
%
\section{Simulation of the process $\ppWN$ }
\begin{figure}[htb]
\begin{center}
\leavevmode
\epsffile{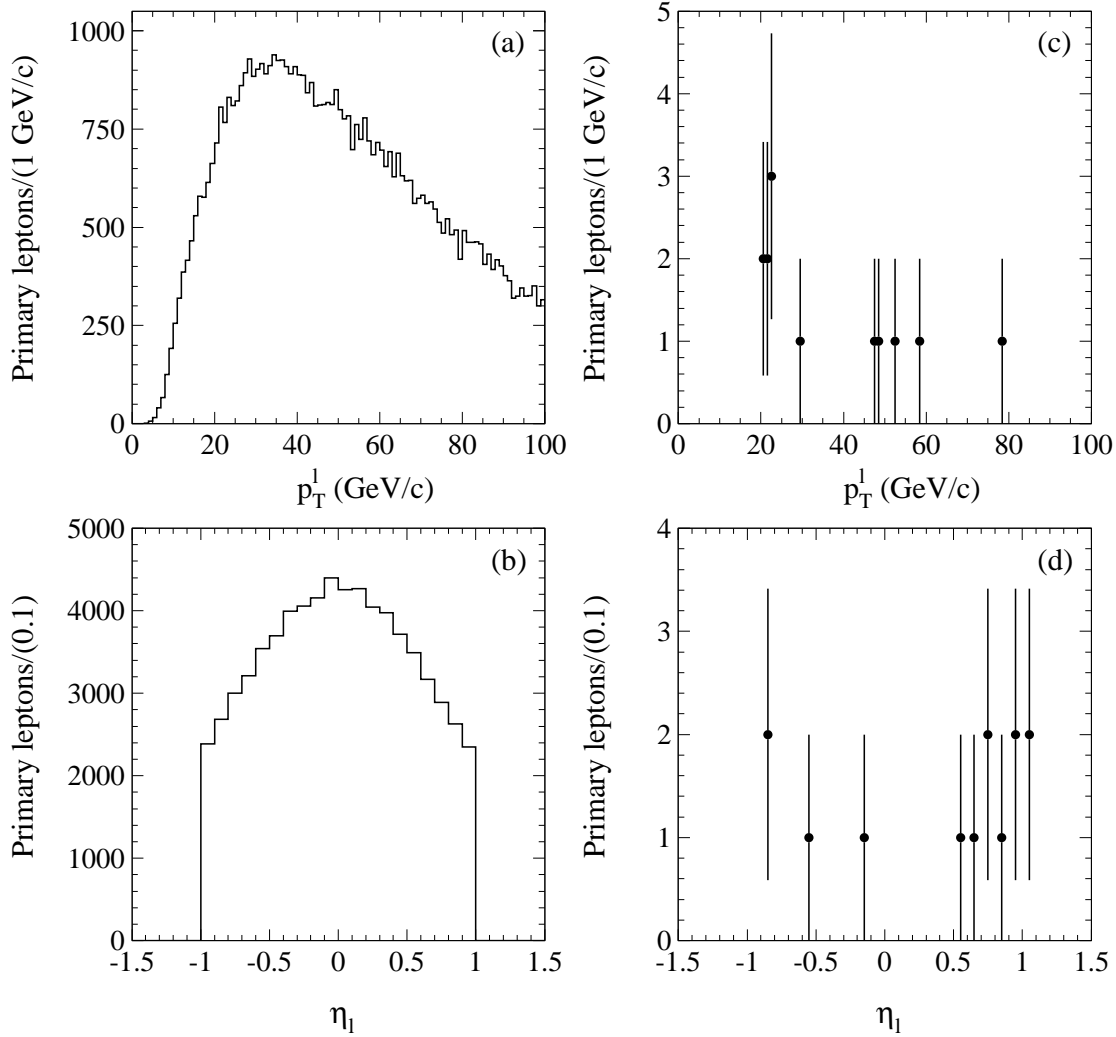}
\caption[]{ Transverse momentum (a) and pseudo-rapidity (b) distributions 
            of the primary leptons resulting from the decay of $\W$
            bosons produced in association with a Higgs  of mass 220 $\gevcc$; 
            (c) and (d) are the analogous distributions for the data.}
\label{fig:fig_11.0}
\end{center}
\end{figure}

\begin{figure}[htb]
\begin{center}
\leavevmode
\epsffile{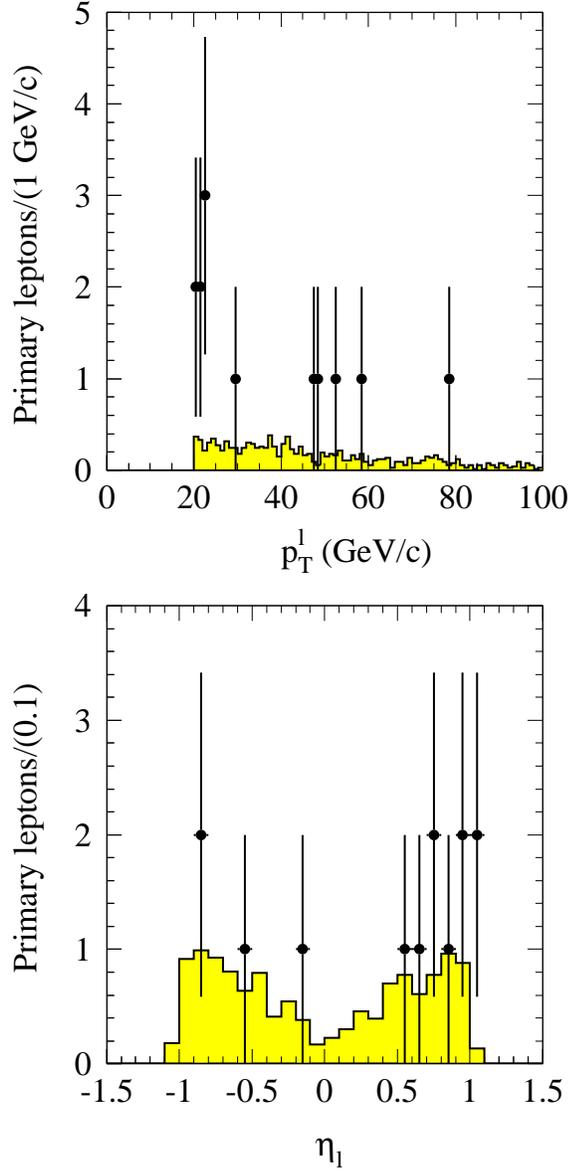}
\caption[]{ Transverse momentum and pseudo-rapidity distributions of the
            primary lepton. The data ($\bullet$) are compared to a
            $q_i \bar{q}_j \rightarrow l \MET N $ simulation
            in which the angular distribution of the primary lepton
            is weighted with the function $\cos^{4} \theta_l$ (see text).}
\label{fig:fig_11.1}
\end{center}
\end{figure}

 We generate the process $q_i \bar{q}_j \rightarrow \W H $ with 
 $m_H$= 220 $\gevcc$ using  option 2605 of {\sc herwig} and the MRS(G) set
 of structure functions~\cite{mrsg}. We identify the Higgs boson with the 
 state $N$ and we force it to decay to $\tilde{b}\bar{b}$. As shown in 
 Figure~\ref{fig:fig_11.0}, the distributions of the pseudo-rapidity and 
 transverse momentum of  primary leptons in this simulation and in the data
 are quite different. We attempt to model the data by considering the process 
 $q_i \bar{q}_j \rightarrow \W N $ as a $2 \rightarrow 3 $ hard scattering
 ($q_i \bar{q}_j \rightarrow l^+ \MET N $, with subsequent decay
 $N \rightarrow \tilde{b} \bar{b}$). For the sake of simplicity,
 we ignore the polarization of the outcoming $W$ calculated by
 {\sc herwig} and decay the $\W$ boson into $l \nu$ according to two-body phase
 space. Then, in order to model the observed  pseudo-rapidity distribution
 of the primary leptons, we weight their distribution in the $q_i \bar{q}_j$
 rest-frame with the function $z^{4}=\cos^{4} \theta_l$,
 where $\theta_l$ is the polar angle of the primary lepton with respect
 to the $q_i$-direction. As shown in Figure~\ref{fig:fig_11.1}, this
 attempt is quite successful. 

\begin{figure}[htb]
\begin{center}
\leavevmode
\epsffile{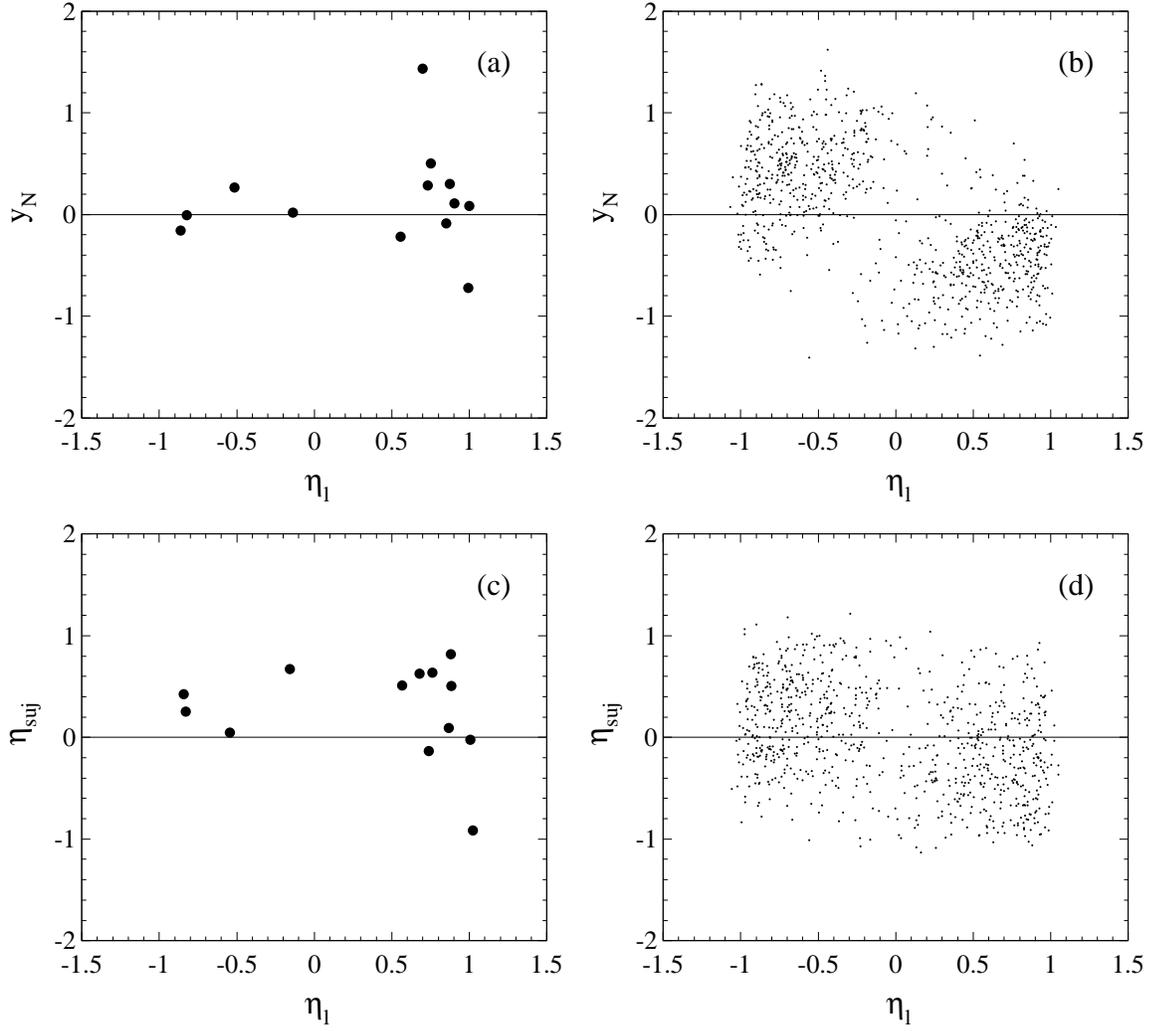}
\caption[]{ Distribution of the rapidity $y_N$ vs the pseudo-rapidity
            $\eta_l$ of the primary lepton in the data (a) and in the modified
            $\W N$ simulation (b). The distribution of the pseudo-rapidity 
            $\eta_{suj}$ of the superjets versus $\eta_l$ is shown in (c) 
            for the data and in (d) for the simulation.}
\label{fig:fig_11.2}
\end{center}
\end{figure}

\begin{figure}[htb]
\begin{center}
\leavevmode
\epsffile{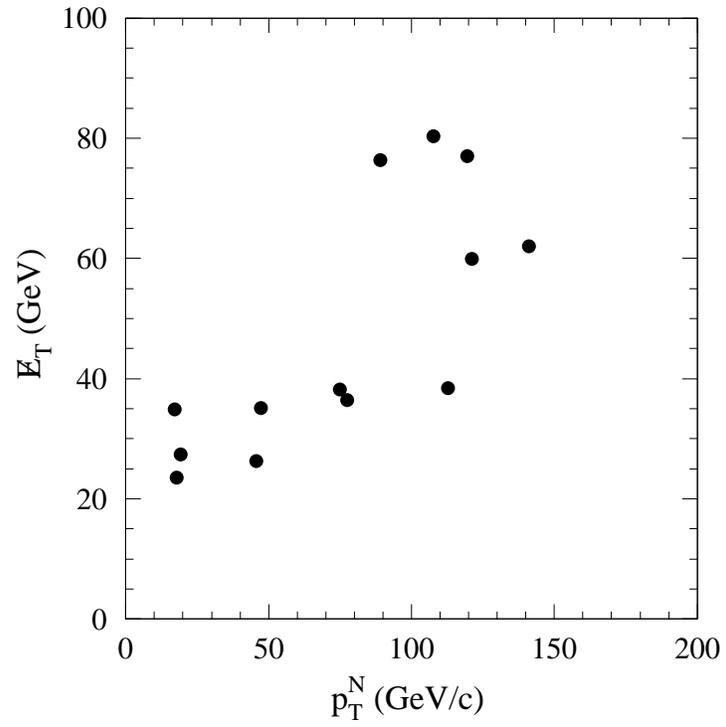}
\caption[]{ Distribution of $\MET$ vs $p_T^N$ in the 13 events with a superjet.}
\label{fig:fig_11.2pri}
\end{center}
\end{figure}

 Using this simulation, we compare additional kinematical quantities to the
 data. Figure~\ref{fig:fig_11.2} shows the correlation between the 
 pseudo-rapidity of the primary lepton and the rapidity of the di-jet 
 system $N$. In the simulation, in which the $\W$ boson recoils against the 
 state $N$, primary leptons tend to have rapidities opposite to the $N$ 
 direction. In the data, the $N$ rapidity is quite central, 
 independent of the primary lepton pseudo-rapidity.
 In the data, primary leptons are produced at large rapidities while
 the state $N$ is produced quite centrally; therefore, most of the $N$ 
 transverse momentum is balanced by the missing transverse energy 
 (Figure~\ref{fig:fig_11.2pri}).

\begin{figure}[htb]
\begin{center}
\leavevmode
\epsffile{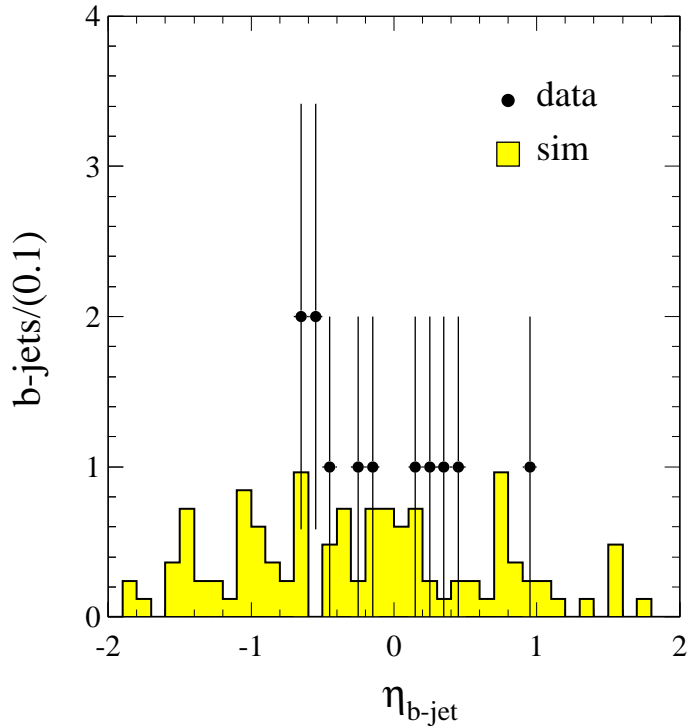}
\caption[]{ Pseudo-rapidity distributions of $b$-jets in the data ($\bullet$)
            and in the $\W N$ simulation (shaded histogram).}
\label{fig:fig_11.2sec}
\end{center}
\end{figure}

 Pseudo-rapidity distributions of the $b$-jets in the data and in the $\W N$
 simulation are shown in Figure~\ref{fig:fig_11.2sec}. Differently
 from the small sample of data, simulated $b$-jets  produced by the decay of 
 a massive state $N$ tend to fill quite uniformly the $\eta$-region covered 
 by the detector.
\newpage
\clearpage
\newpage
%
\vspace*{-2.0cm}
\vspace*{-2.0cm}
%
\section{Effective Lagrangian approach }
 The matrix element used in Sec.~III has been derived  using an effective 
 Lagrangian approach to incorporate the salient features of the data. 
 The production of events with a superjet is modeled with the  
 $2 \rightarrow 4 $ hard scattering process
\begin{eqnarray}
 u(p_1) + \overline{d}(p_2) \rightarrow e^+(p_l) + \nu_s(k) + \bar{b}(p_b)
+ b_s(p_s)
\end{eqnarray}
 where $b_s$ and $\nu_s$ are a scalar quark and a  scalar neutrino.
 A scalar neutrino is used in the final state to model the measured
 missing energy for no compelling reason other than the fact that
 we have already identified superjets with a scalar quark.

 One mechanism capable of producing a $z^4$ angular distribution for the 
 primary leptons in the final state is a high derivative coupling to the 
 lepton and  quark fields. As a first step, using the scalar fields $\phi$
 of the $\nu_s$ and $\tilde{\phi}$ of the $b_s$, the following dimension-1 
 and spin-2 fields are constructed:
$$
\phi_{\mu\nu}(x) = (1/\Lambda)^2\partial_\mu\partial_\nu \phi(x) \mbox{~and~}
\ \tilde{\phi}_{\mu\nu} = (1/\Lambda)^2\partial_\mu\partial_\nu
\tilde{\phi}(x)
$$
 where $\Lambda$ sets the mass scale of the interaction.
 In analogy, from the lepton and quark spinors one constructs
 the following dimension-3/2 and spin-5/2 fields:
$$
\psi_l^{\mu\nu}(x) = (1/\Lambda)^2\partial^\mu\partial^\nu \psi_l(x) \mbox{;~}
\psi_b^{\mu\nu}(x) = (1/\Lambda)^2\partial^\mu\partial^\nu
\psi_b(x)
$$ 
and
\begin{eqnarray}
\chi_q^{\mu\nu}(x) = (1/\Lambda)^2\partial^\mu\partial^\nu
 [ \frac {1+s_q}{2} \psi_u(x) + \frac {1-s_q}{2} \psi_d(x)]  \nonumber \\
\xi_q^{\mu\nu}(x) = (1/\Lambda)^2\partial^\mu\partial^\nu
 [ \frac {1-s_q}{2} \psi_u(x) + \frac {1+s_q}{2} \psi_d(x)] \nonumber 
\end{eqnarray}
 where $s_q$ is the sign of the initial state ($u+\bar{d}$) charge.
 The peculiar forms $\chi_q^{\mu\nu}$ and $\xi_q^{\mu\nu}$, used to account
 for the fields of the initial state partons, are an attempt to also model 
 the asymmetries observed in the rapidity distributions of the data.

 These fields are used to write the following high-derivative, but 
 point-like, effective Lagrangian:
\begin{eqnarray}
{\cal L}(x) = ({{f}\over{\Lambda^{10}}})\partial_\lambda
\{\phi^\dagger_{\alpha\beta}(x) \tilde{\phi}^{'}_{\delta\phi}(x)
{\psi^"}_b^{\tau\sigma}(x)\}\gamma^\rho\partial^\lambda
\{\stackrel{\leftrightarrow \;\;}{\partial^\omega}
\psi_l^{\mu\nu}(x)\stackrel{\leftrightarrow}{\partial}_\omega
[(\partial_\tau\bar{\xi}_q^{\delta\phi}(x))
\gamma_\rho \stackrel{\leftrightarrow}{\partial_\mu}
\stackrel{\leftrightarrow}{\partial_\nu}
(\partial_\sigma\chi_q^{\alpha\beta}(x))]\} \nonumber
\end{eqnarray}
where
$ A \stackrel {\leftrightarrow}{\partial}_{\nu} B =
 A (\partial_{\nu} B) - (\partial_{\nu} A) B $, and $\psi^{"}$ is the
 transpose of  $\psi^{'}$, the ``charge-conjugate" of $\psi$.
 To avoid conflict with unitarity in the high energy limit, we introduce a
 form factor to compensate for the twenty derivatives used 
 above~\footnote{Since we use derivative couplings to fundamental fields, it is
 natural to assume that the hypothetical object involved in
 the $s$-channel exchange is composite.}. 
 However, the effective Lagrangian acquires a reasonable $s$-behaviour 
 only after the inclusion of an additional $s$-channel propagator.
 The squared amplitude for the process (C1) (and its charge conjugate), 
 averaged over initial quark colors, then reads
\begin{eqnarray}
|{\cal M}|^2& =& \left({{f}\over{\Lambda^2}}\right)^2
\frac {\hat{s}^{13}} { (\hat{s} + \Lambda_1^2)^{20}}
\frac { (E_b \;E_l)^5  (k \; E_s)^4}
{ (\hat{s}-M^2)^2 +M^2\Gamma^2} [1- \cos\vartheta_l\; \cos\vartheta_b]
 \times \\ \nonumber
 & &  (1- \frac  {2 E_l} {\sqrt{\hat{s}}})^2 (1- \frac{4 E_l}{\sqrt{\hat{s}}})^2
\left[\cos\vartheta_l \;sin\vartheta_b\;({{1 - \cos\vartheta_k}\over{2}})
(\frac {1 + \cos\vartheta_s} {2} )\right]^4
\end{eqnarray}
 In the kinematics of Eq.~(C2), all six particles in the initial
 and final states are taken to be massless. The energies and polar angles
 [ $(E_l,\vartheta_l)$ for the lepton, $(E_b, \vartheta_b)$ for the b-quark,
 $(k, \vartheta_k)$ for the scalar neutrino, and $( E_s, \vartheta_s)$ for the
 scalar b-quark] are defined in the $u \overline{d}$ center of mass system,
 and $u(p_1)$ is taken to be the in the $z$-direction. Such an effective 
 Lagrangian has been constructed by iterations in which we compared data and
 simulation. We have started using a dimension-6 Lagrangian and we kept 
 increasing the Lagrangian dimension till we obtained an acceptable 
 description of the data.

 The process (C1) has been kindly implemented into the {\sc herwig}
 generator~\cite{moretti}. In the calculation of the matrix element (C2),
 we use $f=1 $ and $\Lambda_1 = 0$ GeV. In order to approximately reproduce the 
 distribution of the invariant mass of the final state in the data, we use 
 $M \simeq$ 350 $\gevcc$ and $\Gamma \simeq$ 5 $\gevcc$
 \footnote{Widths $\Gamma$ as large as 50 $\gevcc$ and masses $M$ in the 
 range of $300-400$ $\gevcc$ also provide quite similar results.}.
 The cross section value of 5 pb derived in Sec.~III corresponds
 to an interaction scale $\Lambda $ of approximately 10 MeV.

\begin{figure}[htb]
\begin{center}
\leavevmode
\epsfxsize \textwidth
\epsffile{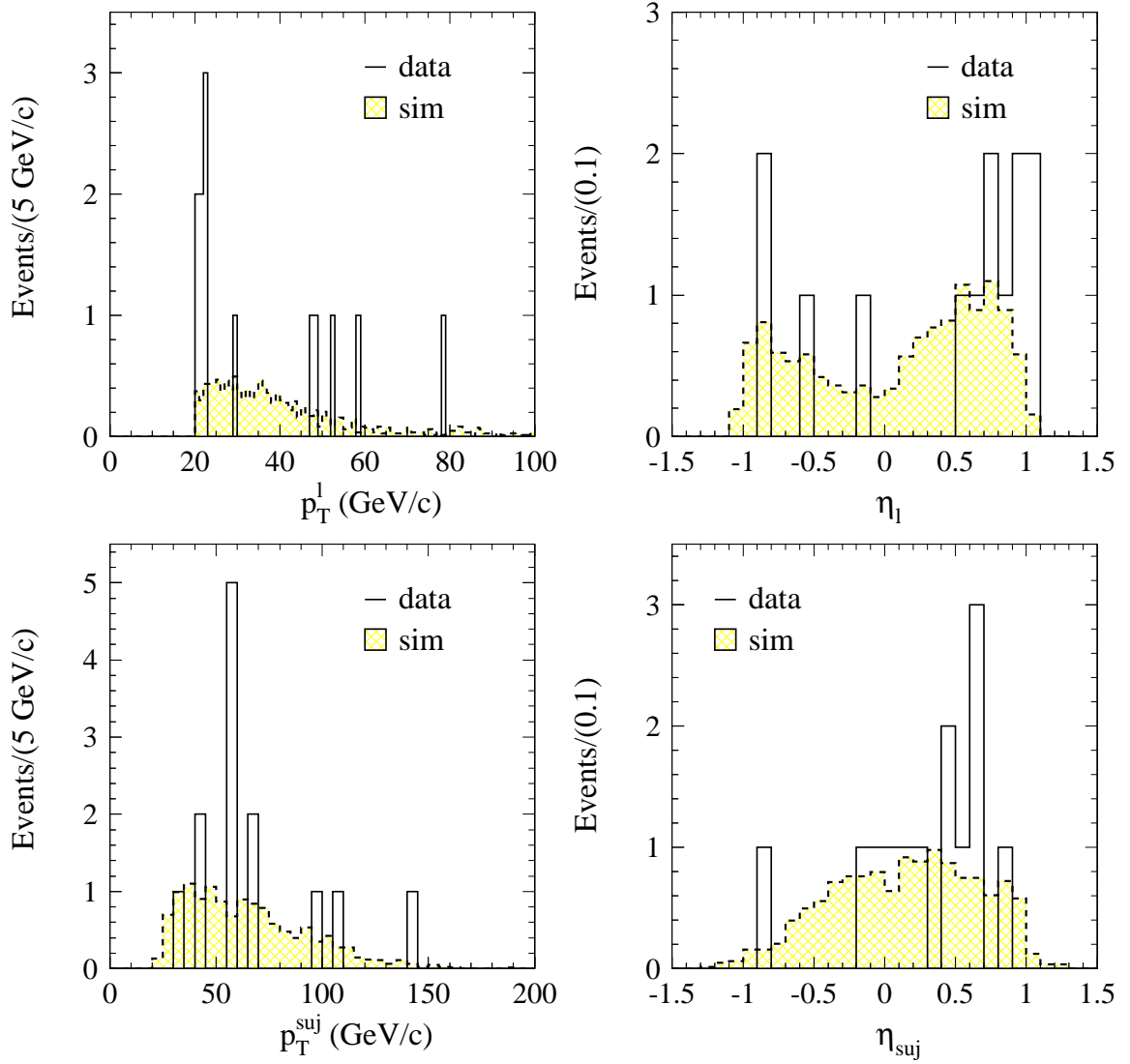}
\caption[]{Transverse momentum and pseudo-rapidity distributions
           of primary leptons and superjets.}
\label{fig:fig_11.3}
\end{center}
\end{figure}

\begin{figure}[htb]
\begin{center}
\leavevmode
\epsfxsize \textwidth
\epsffile{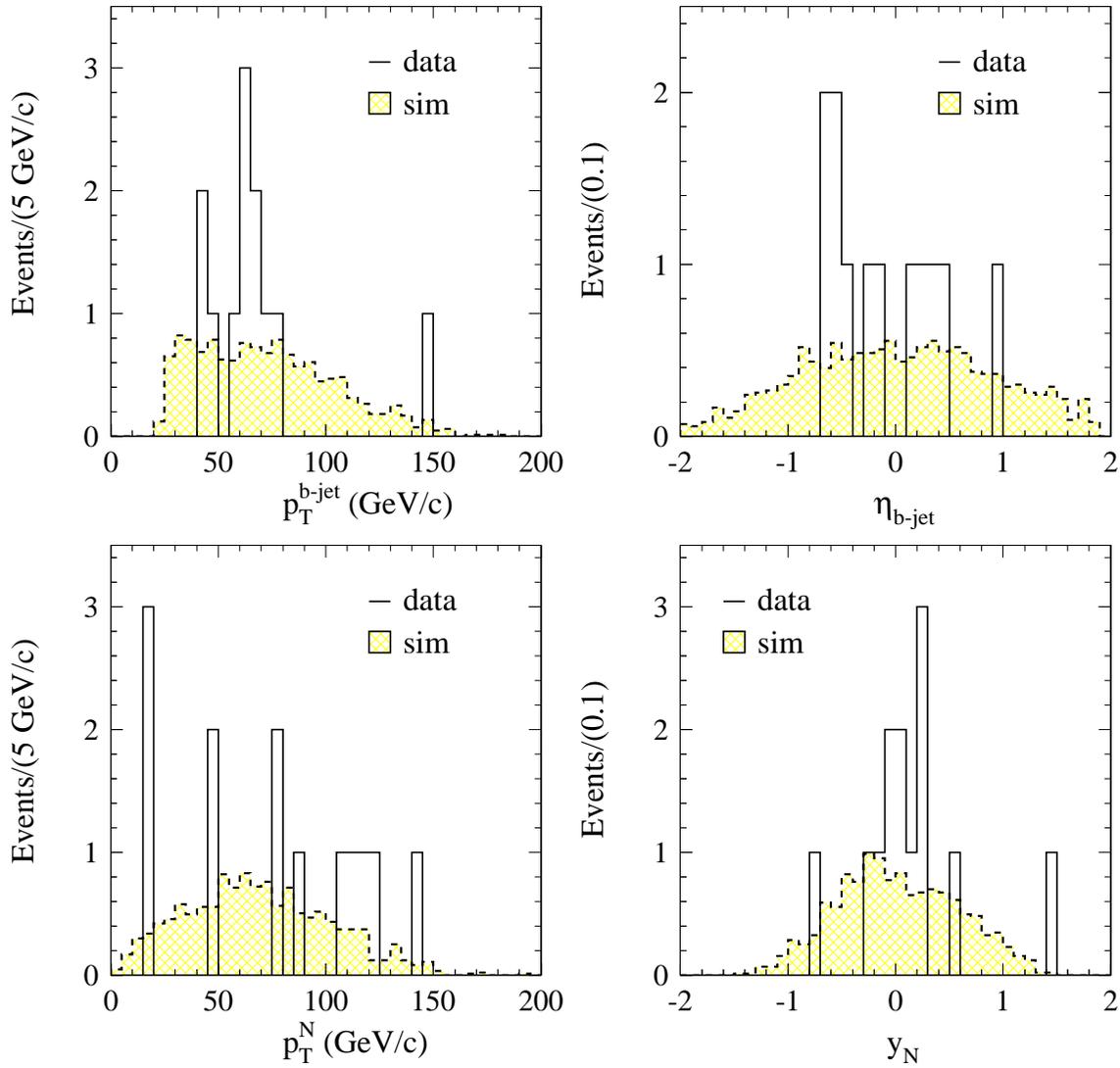}
\caption[]{Transverse momentum and rapidity distributions of $b$-jets and
           of the system $N$ consisting of the $b$-jet and the superjet.}
\label{fig:fig_11.4}
\end{center}
\end{figure}

\begin{figure}[htb]
\begin{center}
\leavevmode
\epsfxsize \textwidth
\epsffile{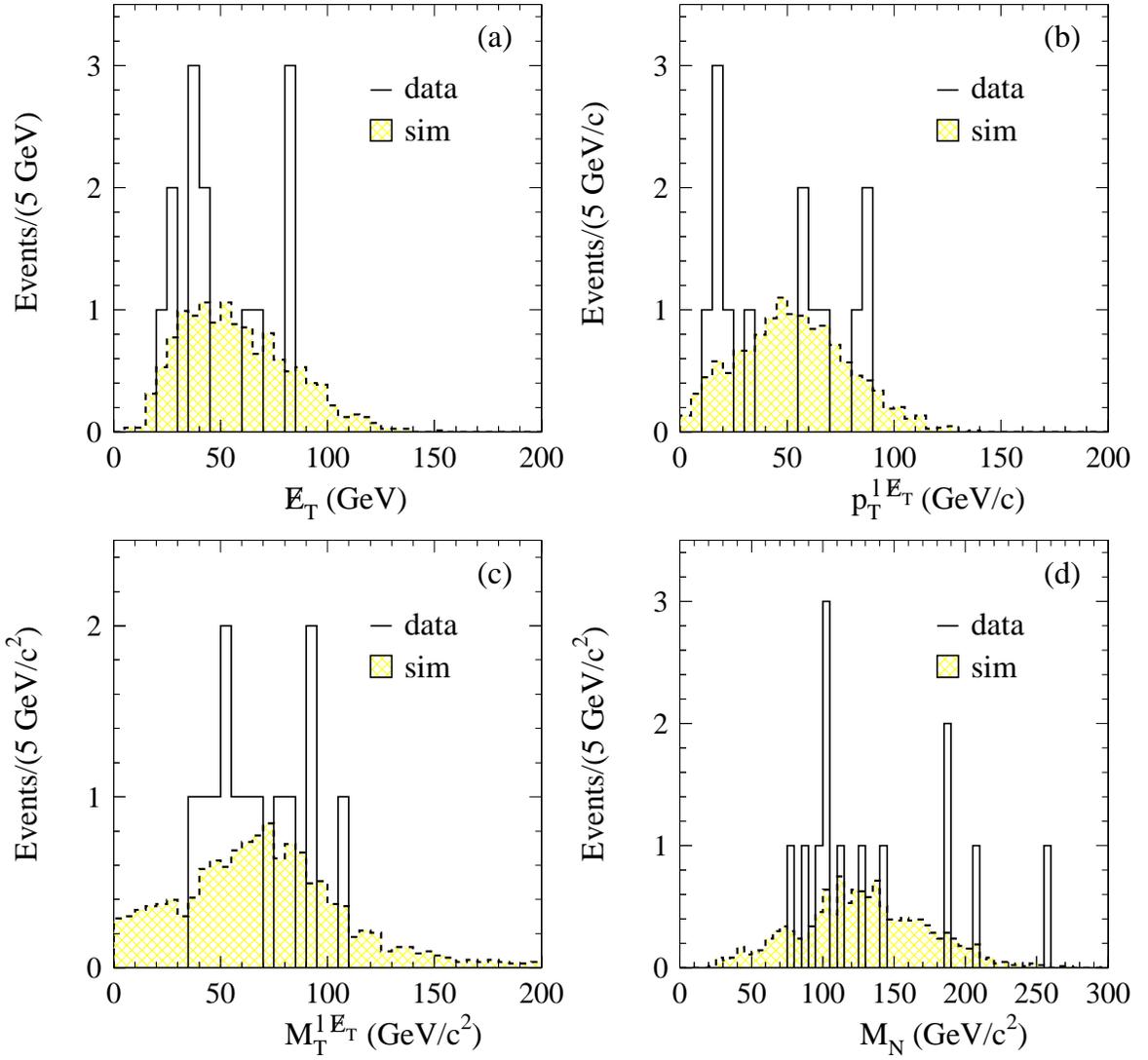}
\caption[]{ Distributions of: $\met$ (a); the transverse momentum of the
            system $l\met$ composed by the primary lepton and the transverse
            missing energy (b); the transverse mass of the system $l\met$ (c);
            and the invariant mass of the system $N$ consisting of the $b$-jet
            and the superjet (d).}
\label{fig:fig_11.5}
\end{center}
\end{figure}

 Figures~\ref{fig:fig_11.3} to~\ref{fig:fig_11.5} compare pseudo-rapidity,
 transverse momentum and invariant mass distributions in the data and the
 simulation normalized to the same number of events. Data and simulation
 are in qualitatively good agreement.
%
\clearpage
%
%
\clearpage

\begin{thebibliography}{99}
\label{bibliography}
\bibitem{anomal}
D.~Acosta {\it et al.,} hep-ex/0109012.
\bibitem{comb}
G.~Apollinari {\it et al.,} hep-ex/0109019.
\bibitem{jpb}   
D.~Buskulic {\it et al.}, Phys.~Lett.~{\bf B313}, 535 (1993).
\bibitem{ttsec} 
T.~Affolder {\it et al.,} Phys.~Rev.~{\bf D64}, 032002 (2001).
\bibitem{mssm-can}
H. E. Haber and G. L. Kane, Phys.~Rep.~{\bf 117C}, 76 (1985).
\bibitem{nappi}
C. Nappi, Phys.~Rev.~{\bf D25}, 84 (1982).
\bibitem{carena}
M.~Carena {\it et al.,} Phys.~Rev.~Lett.~{\bf 86}, 4463 (2001).
\bibitem{cleo-sb}
V.~Savinov {\it et al.,} Phys.~Rev.~{\bf D63}, 051101 (2001).
\bibitem{limit}
 A mass $m_{\tilde{\nu}}$  as large as 10 $\gevcc$ can be
 approximately excluded. In such a case, the energy of the
 scalar quark necessary to produce the observed transverse
 energy of the superjets would be about 200 GeV and, in contrast 
 with the data, the observed missing transverse energy would 
 always be pointing in the direction of the superjet.
\bibitem{tsig-pred}
F.~Bonciani {\it et al.,} Nucl.~Phys.~{\bf B529}, 450 (1998);
E. Berger and H. Contopanagos, Phys. Rev.~{\bf D54}, 3035 (1996);
S. Catani {\it et al.,} Phys.~Lett.~{\bf B378}, 329 (1996);
E. Laenen {\it et al.,} Phys.~Lett.~{\bf B321}, 254 (1994);
P. Nason {\it et al.,} Nucl.~Phys.~{\bf B303}, 607 (1998).
\bibitem{d0}
S.~Abachi {\it et al.,} Phys.~Rev.~Lett.~{\bf 79}, 12003 (1997).
\bibitem{herwig}
  G.~Marchesini and B.~R.~Webber, Nucl.~Phys.~{\bf B310}, 461
 (1988); G.~Marchesini {\it et al.,} Comput.~Phys.~Comm.~{\bf 67}, 465 (1992).
\bibitem{cleo}  
  P.~Avery, K.~Read, G.~Trahern, Cornell Internal Note
               CSN-212, March 25, 1985 (unpublished). We use Version
               9\_1 of the CLEO simulation and our own lifetime database.
\bibitem{sq-koba}
K. Hikasa and M. Kobayashi, Phys.~Rev.~{\bf D36}, 724 (1987);
J.~F.~Gunion and H.~E.~Haber, Nucl.~Phys.~{\bf B272}, 1 (1986).
\bibitem{mrsg} A.~D.~Martin, R.~G.~Roberts and W.~J.~Stirling,
               Phys.~Lett.~{\bf B354}, 155 (1995).
\bibitem{moretti}
 The process (C1) has been implemented into
 the {\sc herwig} simulation by S. Moretti (routine {\sc hwgrvt}).

\end{thebibliography}
\end{document}